\documentclass[journal,10pt]{IEEEtran}
\usepackage{epsfig}
\usepackage{booktabs}
\usepackage{epstopdf}
\usepackage{color}
\usepackage{amssymb}
\usepackage{mathtools}

\begin{document}

\title{Secure mmWave Communications in Cognitive Radio Networks}

\author{Hui Zhao,~\IEEEmembership{Student Member,~IEEE}, Jiayi Zhang,~\IEEEmembership{Member,~IEEE}, Liang Yang,~\IEEEmembership{Member,~IEEE},\\ Gaofeng Pan,~\IEEEmembership{Member,~IEEE},  and Mohamed-Slim Alouini,~\IEEEmembership{Fellow,~IEEE} \vspace{-1cm}

\thanks{Manuscript received February 25, 2019; accepted April 6, 2019. The associate editor coordinating the review of this letter and approving it for publication was K. Adachi. (\emph{Corresponding author: Gaofeng Pan}.)}

\thanks{H. Zhao,  and M.-S. Alouini are with the Computer, Electrical, and Mathematical Science
and Engineering Division, King Abdullah University of Science and Technology, Thuwal 23955-6900, Saudi Arabia (email: hui.zhao@kaust.edu.sa; slim.alouini@kaust.edu.sa).}

\thanks{J. Zhang  is with the School of Electronic and Information
Engineering, Beijing Jiaotong University, Beijing 100044, China (e-mail:
jiayizhang@bjtu.edu.cn).}
\thanks{L. Yang is with  the College of Computer Science and Electronic Engineering, Hunan University, Changsha
410082, China (e-mail: liangyang.guangzhou@gmail.com).}

\thanks{G. Pan is with the School of Information and Electronics, Beijing Institute of Technology, Beijing 100081, China (email: Gaofeng.Pan.CN@ieee.org).}
\thanks{Digital Object Identifier 10.1109/LWC.2019.2910530}

% <-this % stops a space
}

\markboth{IEEE Wireless Communications Letters,~Vol.~XX, No.~XX, XXX~2019}
{}
%{Shell \MakeLowercase{\textit{et al.}}: Bare Demo of IEEEtran.cls for Journals}

\maketitle

\begin{abstract}
In this letter, the secrecy performance in cognitive radio networks (CRNs) over fluctuating two-ray (FTR) channels, which is used to model the milimeter wave channel, is investigated in terms of the secrecy outage probability (SOP). Specifically, we consider the case where a source ($S$) transmits confidential messages to a destination ($D$), and an eavesdropper wants to wiretap the information from $S$ to $D$. In a CRN framework, we assume that the primary user shares its spectrum with $S$, where $S$ adopts the underlay strategy to control its transmit power without impairing the quality of service of the primary user. After some mathematical manipulations, an exact analytical expression for the SOP is derived. In order to get physical and technical insights into the effect of the channel parameters on the SOP, we derive an asymptotic formula for the SOP  in the high signal-to-noise ratio region of the $S-D$ link. We finally show some selected  Monte-Carlo simulation results to validate the correctness of our derived analytical expressions.
\end{abstract}


\IEEEpeerreviewmaketitle

\section{Introduction}
Two major technologies to enhance the spectrum efficiency are cognitive radio networks (CRNs)  and millimeter wave (mmWave) communications \cite{Sboui}-\cite{Mezzavilla}, where the first one allows the primary users to share the spectrum with secondary users without impairing the quality of service (QoS) of primary users in CRNs by using some protection strategies \cite{Sboui}, and another provides large available bandwidth at mmWave frequencies \cite{Mezzavilla}.
Among common protection strategies in CRNs, the underlay scheme is the  simplest one, because  secondary users only adjust their transmit power without exceeding  a certain interference threshold at primary users, which is easy to perform in practical CRNs \cite{Ahmad}. To this end, the underlay scheme has been  received an increasing attention \cite{Sboui}-\cite{Lee}.

One major obstacle to realize mmWave communications is to characterize the mmWave channel, especially the random fluctuation suffered by the received signal, which has been properly solved by \cite{Goldsmith} where the fluctuating two-ray (FTR) channel model was proposed. In the FTR channel model, specular waves randomly fluctuate, rather than a constant amplitude in the two-wave with diffuse power (TWDP) channel model, which means that the FTR model is a natural generalization of TWDP model. In fact, the FTR model can reduce to many traditional fading channels, such as Rician and Nakagimi-$m$ channels, where the relative parameter settings of FTR model are shown in the Table I of \cite{Goldsmith}. Recently, \cite{Zhang1,huige2} has extended \cite{Goldsmith} to a more generalized FTR channel model allowing any positive value of $m$.

Physical layer security is a common topic in wireless communications because of the open access \cite{Bloch}-\cite{Liang}.
 For example, the secrecy outage probability (SOP) of point-to-point digital communications, i.e., the typical three-node Wyner's model in \cite{Bloch} over FTR channels,  was investigated in \cite{Zhang2}.
However, to best of authors' knowledge, there is no related work on physical layer security in CRNs over FTR fading channels, and the secrecy analysis is an important issue in CRNs \cite{Tang}-\cite{huige1}, because the frequency band shared among primary and secondary users results in a higher interception probability both in the primary and secondary networks.

To fill this gap, we investigate the physical layer security in CRNs over FTR channels in terms of the SOP, and derive  the analytical expression for the exact SOP. The asymptotic SOP (ASOP) has been also proposed with simple functions to cut down the computation complexity in the high signal-to-noise ratio (SNR) region. Moreover, the secrecy diversity order and secrecy array gain of ASOP are presented to reveal the physical insights of channel parameters on the security performance of CRNs.

\section{System Model}
There is a source ($S$) transmitting signal to a destination ($D$) in a secondary network, where a primary user ($P$) shares the spectrum with $S$. However, an eavesdropper ($E$) wants to overhear the information from $S$ to $D$. $h_p$, $h_d$, and $h_e$ are the channel power gains of the $S-P$, $S-D$, and $S-E$ links, respectively. In the underlay scheme, the transmit power ($P_t$) of $S$ should be less than a certain threshold ($I_{th}$) to guarantee the QoS of $P$, i.e.,
\begin{align}
{P_t} = \min \left\{ {\frac{{{I_{th}}}}{{{h_p}}},{P_M}} \right\} = \mathbb{I}\left\{ {{h_p} \ge \rho } \right\}\frac{{{I_{th}}}}{{{h_p}}} + \mathbb{I}\left\{ {{h_p} < \rho } \right\}{P_M},
\end{align}
where $\rho  = {{{I_{th}}} \mathord{\left/{\vphantom {{{I_{th}}} {{P_M}}}} \right.\kern-\nulldelimiterspace} {{P_M}}}$,  $P_M$ is the maximal transmit power of $S$,  and $\mathbb{I}\{\cdot\}$ denotes the indicator function, i.e., $\mathbb{I}\{\mathcal{A}\}$ is unity for $\mathcal{A}$ true and zero otherwise. It is worth noting that (1) reveals the main difference from the work in \cite{Zhang2} where the transmitter only uses a fixed transmit power, i.e., $P_M$, for communications with the legitimate receiver. If $I_{th} \to \infty$ (or equivalently $\rho \to \infty$), our system will become the typical three-node Wyner's model investigated in \cite{Zhang2}.

We assume that all links follow independent FTR fading.
The probability density function (PDF) and cumulative density function (CDF) of $h_t$ $(t \in \{p,d,e\})$ are given by \cite{Zhang1}
\begin{align}
{f_{{h_t}}}\left( x \right) &= \frac{{m_t^{{m_t}}}}{{\Gamma \left( {{m_t}} \right)}}\sum\limits_{{j_t} = 0}^\infty  {\frac{{K_t^{{j_t}}{d_{{j_t}}}{x^{{j_t}}}}}{{{j_t}!{j_t}!{{\left( {2\sigma _t^2} \right)}^{{j_t} + 1}}}}} \exp \left( { - \frac{x}{{2\sigma _t^2}}} \right), \\
{F_{{h_t}}}\left( x \right) &= 1 - \frac{{m_t^{{m_t}}}}{{\Gamma \left( {{m_t}} \right)}}\sum\limits_{{j_t} = 0}^\infty  {\frac{{K_t^{{j_t}}{d_{{j_t}}}}}{{{j_t}!\exp \left( {\frac{x}{{2\sigma _t^2}}} \right)}}} {\rm{ }}\sum\limits_{{n_t} = 0}^{{j_t}} {\frac{{{{\left( {{x \mathord{\left/
 {\vphantom {x {\left( {2\sigma _t^2} \right)}}} \right.
 \kern-\nulldelimiterspace} {\left( {2\sigma _t^2} \right)}}} \right)}^{{n_t}}}}}{{{n_t}!}}},
\end{align}
respectively, where $\Gamma(\cdot)$ denotes the Gamma function \cite{Gradshteyn}, $m_t$ is the parameter of Gamma distribution with unit mean, $K_{t}$ is the average power ratio of the dominant waves and remaining diffuse multipath, $\sigma_t^2$ is the variance of the real (or imaginary) diffuse component, and the definition of $d_{j_t}$ is
\begin{align}
{d_{{j_t}}} =& \sum\limits_{k = 0}^{{j_t}} {j_t \choose k} {\left( {\frac{{{\Delta _t}}}{2}} \right)^k}\sum\limits_{l = 0}^k {k \choose l}\Gamma \left( {{j_t} + {m_t} + 2l - k} \right) \notag\\
&\frac{{{e^{\frac{{\pi \left( {2l - k} \right)i}}{2}}}P_{{j_t} + {m_t} - 1}^{k - 2l}\left( {\frac{{{m_t} + {K_t}}}{{\sqrt {{{\left( {{m_t} + {K_t}} \right)}^2} - {{\left( {{K_t}{\Delta _t}} \right)}^2}} }}} \right)}}{{{{\left( {\sqrt {{{\left( {{m_t} + {K_t}} \right)}^2} - {{\left( {{K_t}{\Delta _t}} \right)}^2}} } \right)}^{{j_t} + {m_t}}}}},
\end{align}
where $\Delta_t \in [0,1]$ is to characterize the relation of two dominant wave powers, $i$ is the imaginary unit, and $P(\cdot)$ denotes the Legendre function of the first kind \cite{Gradshteyn}.
From (5) in \cite{Zhang1}, the expectation of $h_t$ is $\mu_t=\mathbb{E}\{h_t\}=2\sigma_t^2 (1+K_t)$, where $\mathbb{E}\{\cdot\}$ denotes the expectation operator.

The equivalent SNRs at $D$ and $E$ can be expressed as\footnote{We assume that there is no interference from the primary network, due to the fact that the primary transmitter is far from both $D$ and $E$ (or the primary transmitter employs the random Gaussian codebook, and the interference from the primary user at $D$ and $E$ can be represented by noise) \cite{Lee}.}
\begin{align}\label{gamma_d}
&{\gamma _d}= \mathbb{I}\left\{ {{h_p} \ge \rho } \right\}\frac{{{I_{th}}{h_d}}}{{{N_0}{h_p}}} + \mathbb{I}\left\{ {{h_p} < \rho } \right\}\frac{{{P_M}{h_d}}}{{{N_0}}}, \\
&{\gamma _e}= \mathbb{I}\left\{ {{h_p} \ge \rho } \right\}\frac{{{I_{th}}{h_e}}}{{{N_0}{h_p}}} + \mathbb{I}\left\{ {{h_p} < \rho } \right\}\frac{{{P_M}{h_e}}}{{{N_0}}},
\end{align}
respectively, where $N_0$ denotes the power of the Gaussian noise at receivers.
\vspace{-0.5cm}
\section{Secrecy Outage Probability}
We assume that $S$ only has the instantaneous channel state information (CSI) of $S-D$ link, and does not know the CSI of  $S-E$ link, and therefore, $S$ has no choice but to transmit signal at a constant rate of confidential information ($R_s$). In this case, perfect security cannot be guaranteed, because the instantaneous secrecy capacity defined in \cite{Bloch},
$
{C_s} = \max \left\{ {{{\log }_2}\left( {1 + {\gamma _d}} \right) - {{\log }_2}\left( {1 + {\gamma _e}} \right),0} \right\},
$ cannot be always greater than the target secrecy rate ($R_s$).
The SOP is to capture the secrecy outage performance, the probability that $R_s$ is greater than the secrecy capacity \cite{Bloch}, i.e.,
\begin{align}\label{SOP_def}
&{\rm{SOP}}=\Pr\{C_s \le R_s\}= \Pr \left\{ {{\gamma _d} \le \lambda  - 1 + \lambda {\gamma _e}} \right\},
\end{align}
where $\lambda=2^{R_s}$.
By substituting \eqref{gamma_d} and (6) into \eqref{SOP_def}, the SOP is written as
\begin{align}
&{\rm{SOP}} = \int_0^\infty  {\Pr \left\{ {\mathbb{I}\left\{ {{h_p} \ge \rho } \right\}\frac{{{I_{th}}{h_d}}}{{{N_0}{h_p}}} + \mathbb{I}\left\{ {{h_p} < \rho } \right\}\frac{{{P_M}{h_d}}}{{{N_0}}}} \right.}  \notag\\
&\hspace{0.5cm}\left. {\left. { \le \lambda  - 1 + \mathbb{I}\left\{ {{h_p} \ge \rho } \right\}\frac{{\lambda {I_{th}}{h_e}}}{{{N_0}{h_p}}} + \mathbb{I}\left\{ {{h_p} < \rho } \right\}\frac{{\lambda {P_M}{h_e}}}{{{N_0}}}} \right|{h_p}} \right\}\notag\\
&\hspace{0.5cm}{f_{{h_p}}}\left( {{h_p}} \right)d{h_p}.
\end{align}
By using the definition of the indicator function $\mathbb{I}\{\cdot\}$, the SOP can be further written as
\begin{align}\label{SOP_int}
&{\rm{SOP }} = \underbrace {{F_{{h_p}}}\left( \rho  \right)\Pr \left\{ {\frac{{{P_M}{h_d}}}{{{N_0}}} \le \lambda  - 1 + \frac{{\lambda {P_M}{h_e}}}{{{N_0}}}} \right\}}_{{\rm{SO}}{{\rm{P}}_{\rm{1}}}} \notag\\
 &+ \underbrace {\int_\rho ^\infty  {\Pr \left\{ {\left. {\frac{{{I_{th}}{h_d}}}{{{N_0}{h_p}}} \le \lambda  - 1 + \frac{{\lambda {I_{th}}{h_e}}}{{{N_0}{h_p}}}} \right|{h_p}} \right\}} {f_{{h_p}}}\left( {{h_p}} \right)d{h_p}}_{{\rm{SO}}{{\rm{P}}_{\rm{2}}}}.
\end{align}
It is obvious that the ${\rm SOP}_1$ is the product of the probability of $h_p <\rho$ and  the SOP in non-CRNs where $S$ transmits signal to $D$ at a fixed transmit power, i.e., $P_M$, where the SOP in non-CRNs has been investigated in \cite{Zhang2},  given by \eqref{SOP1} (refer to Lemma 2 in \cite{Zhang2}), shown on the top of next page, %Lemma 2 in \cite{Zhang2},
\begin{figure*}
\begin{align}\label{SOP1}
\resizebox{.9\hsize}{!}{$
{\rm{SO}}{{\rm{P}}_1} = {F_{{h_p}}}\left( \rho  \right) - \frac{{{F_{{h_p}}}\left( \rho  \right)m_d^{{m_d}}}}{{\Gamma \left( {{m_d}} \right)}}\sum\limits_{{j_d} = 0}^\infty  {\frac{{K_d^{{j_d}}{d_{{j_d}}}}}{{{j_d}!}}} \sum\limits_{{n_d} = 0}^{{j_d}} {\frac{{\exp \left( {\frac{{ - {N_0}\left( {\lambda  - 1} \right)}}{{{P_M}2\sigma _d^2}}} \right)}}{{{n_d}!{{\left( {2\sigma _d^2} \right)}^{{n_d}}}}}}
\sum\limits_{f = 0}^{{n_d}} {n_d \choose f} {\left( {\frac{{{N_0}\left( {\lambda  - 1} \right)}}{{{P_M}}}} \right)^{{n_d} - f}}{\lambda ^f}\mathbb{E}\left\{ {h_e^f\exp \left( {\frac{{ - \lambda {h_e}}}{{2\sigma _d^2}}} \right)} \right\}.
$}
\end{align}
\rule{18cm}{0.01cm}
\end{figure*}
where
\begin{align}\label{Expectation}
&\mathbb{E}\left\{ {h_e^f\exp \left( { - \frac{{\lambda {h_e}}}{{2\sigma _d^2}}} \right)} \right\} = \int_0^\infty  {h_e^f} \exp \left( { - \frac{{\lambda {h_e}}}{{2\sigma _d^2}}} \right){f_{{h_e}}}\left( {{h_e}} \right)d{h_e} \notag\\
& = \frac{{m_e^{{m_e}}}}{{\Gamma \left( {{m_e}} \right)}}\sum\limits_{{j_e} = 0}^\infty  {\frac{{K_e^{{j_e}}{d_{{j_e}}}}}{{{j_e}!{j_e}!{{\left( {2\sigma _e^2} \right)}^{{j_e} + 1}}}}} \frac{{\Gamma \left( {{j_e} + f + 1} \right)}}{{{{\left( {\frac{\lambda }{{2\sigma _d^2}} + \frac{1}{{2\sigma _e^2}}} \right)}^{{j_e} + f + 1}}}}.
\end{align}

We can further write ${\rm SOP}_2$  in the complementary CDF (CCDF) form as
\begin{align}
{\rm{SO}}{{\rm{P}}_{\rm{2}}}=& {\overline F _{{h_p}}}\left( \rho  \right) - \int_\rho ^\infty  {\int_0^\infty  {{{\overline F }_{{h_d}}}\left( {\frac{{\left( {\lambda  - 1} \right){N_0}{h_p}}}{{{I_{th}}}} + \lambda {h_e}} \right)} } \notag\\
&\hspace{0.5cm}{f_{{h_e}}}\left( {{h_e}} \right)d{h_e}{f_{{h_p}}}\left( {{h_p}} \right)d{h_p},
\end{align}
where $\overline F_{h_t}(\cdot)$ ($t \in \{p,d,e\}$) denotes the CCDF of $h_t$.

After some mathematical manipulations, ${\rm SOP}_2$ can be derived as
\begin{align}
{\rm{SO}}{{\rm{P}}_{\rm{2}}} =& {\overline F _{{h_p}}}\left( \rho  \right) - \frac{{m_d^{{m_d}}}}{{\Gamma \left( {{m_d}} \right)}}\sum\limits_{{j_d} = 0}^\infty  {\frac{{K_d^{{j_d}}{d_{{j_d}}}}}{{{j_d}!}}}\sum\limits_{{n_d} = 0}^{{j_d}} {\frac{1}{{{n_d}!}}{{\left( {\frac{1}{{2\sigma _d^2}}} \right)}^{{n_d}}}} \notag\\
& \sum\limits_{f = 0}^{{n_d}}{n_d \choose f} {{{\left( {\frac{{\left( {\lambda  - 1} \right){N_0}}}{{{I_{th}}}}} \right)}^{{n_d} - f}}} {\lambda ^f}\notag\\
&\underbrace {\int_\rho ^\infty  {h_p^{{n_d} - f}\exp \left( { - \frac{{\left( {\lambda  - 1} \right){N_0}{h_p}}}{{2\sigma _d^2{I_{th}}}}} \right){f_{{h_p}}}\left( {{h_p}} \right)d{h_p}} }_{{\mathcal{I}_1}} \notag\\
&\underbrace {\int_0^\infty  {h_e^f\exp \left( { - \frac{{\lambda {h_e}}}{{2\sigma _d^2}}} \right)} {f_{{h_e}}}\left( {{h_e}} \right)d{h_e}}_{{\mathcal{I}_2}},
\end{align}
where
\begin{align}\label{I1}
\resizebox{.9\hsize}{!}{$
{{\cal I}_1} = \frac{{m_p^{{m_p}}}}{{\Gamma \left( {{m_p}} \right)}}\sum\limits_{{j_p} = 0}^\infty  {\frac{{K_p^{{j_p}}{d_{{j_p}}}}}{{{j_p}!{j_p}!{{\left( {2\sigma _p^2} \right)}^{{j_p} + 1}}}}} \frac{{{\rm{ }}\Gamma \left( {{j_p} + {n_d} - f + 1,\frac{{\left( {\lambda  - 1} \right){N_0}\rho }}{{2\sigma _d^2{I_{th}}}} + \frac{\rho }{{2\sigma _p^2}}} \right)}}{{{{\left( {\frac{{\left( {\lambda  - 1} \right){N_0}}}{{2\sigma _d^2{I_{th}}}} + \frac{1}{{2\sigma _p^2}}} \right)}^{{j_p} + {n_d} - f + 1}}}},
$}
\end{align}
and
\begin{align}\label{I2}
{\mathcal{I}_2} = \frac{{m_e^{{m_e}}}}{{\Gamma \left( {{m_e}} \right)}}\sum\limits_{{j_e} = 0}^\infty  {\frac{{K_e^{{j_e}}{d_{{j_e}}}\Gamma \left( {{j_e} + f + 1} \right)}}{{{j_e}!{j_e}!{{\left( {2\sigma _e^2} \right)}^{{j_e} + 1}}}}} \frac{1}{{\left( {\frac{\lambda }{{2\sigma _d^2}} + \frac{1}{{2\sigma _e^2}}} \right)^{ {j_e} +f + 1}}},
\end{align}
where $\Gamma(\cdot,\cdot)$ denotes the upper incomplete Gamma function \cite{Gradshteyn}.
In view of  expressions for ${\rm SOP}_1$ and ${\rm SOP}_2$, the exact  expression for  SOP is derived as \eqref{SOP}, shown on the top of next page, where the expressions for $\mathcal{I}_1$, $\mathcal{I}_2$ and $\mathbb{E}\left\{ {h_e^f\exp \left( { - \frac{{\lambda {h_e}}}{{2\sigma _d^2}}} \right)} \right\}$ can be found in \eqref{I1}, \eqref{I2} and \eqref{Expectation}, respectively.
\begin{figure*}
\begin{align}\label{SOP}
{\rm SOP} =& 1 - \frac{{{F_{{h_p}}}\left( \rho  \right)m_d^{{m_d}}}}{{\Gamma \left( {{m_d}} \right)}}\sum\limits_{{j_d} = 0}^\infty  {\frac{{K_d^{{j_d}}{d_{{j_d}}}}}{{{j_d}!}}} \sum\limits_{{n_d} = 0}^{{j_d}} {\frac{1}{{{n_d}!{{\left( {2\sigma _d^2} \right)}^{{n_d}}}}}} {\exp \left( {\frac{{ - {N_0}\left( {\lambda  - 1} \right)}}{{{P_M}2\sigma _d^2}}} \right)} \sum\limits_{f = 0}^{{n_d}} {n_d \choose f} {\left( {\frac{{{N_0}\left( {\lambda  - 1} \right)}}{{{P_M}}}} \right)^{{n_d} - f}}{\lambda ^f} \notag\\
&\mathbb{E}\left\{ {h_e^f\exp \left( {\frac{{ - \lambda {h_e}}}{{2\sigma _d^2}}} \right)} \right\} - \frac{{m_d^{{m_d}}}}{{\Gamma \left( {{m_d}} \right)}}\sum\limits_{{j_d} = 0}^\infty  {\frac{{K_d^{{j_d}}{d_{{j_d}}}}}{{{j_d}!}}} \sum\limits_{{n_d} = 0}^{{j_d}} {\frac{1}{{{n_d}!}}{{\left( {\frac{1}{{2\sigma _d^2}}} \right)}^{{n_d}}}} \sum\limits_{f = 0}^{{n_d}} {n_d \choose f} {\left( {\frac{{\left( {\lambda  - 1} \right){N_0}}}{{{I_{th}}}}} \right)^{{n_d} - f}}{\lambda ^f}{\mathcal{I}_1}{\mathcal{I}_2}.
\end{align}
\rule{18cm}{0.01cm}
\end{figure*}

\vspace{-0.5cm}
\section{Asymptotic Analysis}
From \eqref{SOP_int}, we can easily see that ${\rm SOP}\approx {\rm SOP}_1$ for $\rho \to \infty$, and ${\rm SOP}\approx {\rm SOP}_2$ for $\rho \to 0$. Therefore, one way to approximate the SOP is
\begin{align}
{\rm SOP}\approx\begin{cases}
\left. {\rm SOP}_1 \right|_{\rho \to \infty}, &\text{if }\rho \gg 0;\\
\left. {\rm SOP}_2 \right|_{\rho=0}, &\text{if }\rho \to 0,
\end{cases}
\end{align}
where $\left. {\rm SOP}_2 \right|_{\rho=0}$ is actually the proposed SOP by \cite{Yuanwei2} without taking the maximal transmit power constraint at the transmitter into account, and ${\left. {{\rm{SO}}{{\rm{P}}_1}} \right|_{\rho  \to \infty }}$ is  the SOP in non-CRNs, i.e., the SOP investigated in \cite{Zhang2}.

To get the secrecy diversity order and secrecy array gain  for the SOP, we analyze the ASOP when $\mu_d \to \infty$ and $\mu_e$ remains constant.
The asymptotic CDF of $h_d$ for $2\sigma _d^2 \gg 0$ is given by (32) in \cite{Goldsmith}
\begin{align}
&F_{{h_d}}^\infty \left( x \right) = \frac{{m_d^{{m_d}}{d_{{j_d} = 0}}x}}{{\Gamma \left( {{m_d}} \right)}}{\left( {2\sigma _d^2} \right)^{ - 1}} + o\left( {{{\left( {2\sigma _d^2} \right)}^{ - 2}}} \right),
\end{align}
where $o\left(  \cdot  \right)$ denotes the higher order term, and $d_{j_d=0}$ is the value of $d_{j_d}$ given $j_d=0$.

Using the asymptotic CDF of $h_d$ and some mathematical manipulations, we can derive the asymptotic ${\rm SOP}_1$ and ${\rm SOP}_2$ as
{\color{red}{
\begin{align}\label{SOP1_asy}
&{\rm{SOP}}_1^\infty  = \frac{{{F_{{h_p}}}\left( \rho  \right)m_d^{{m_d}}{d_{{j_d} = 0}}}}{{\Gamma \left( {{m_d}} \right)}}\left( {\frac{(\lambda-1)N_0}{P_M} + \lambda {\mu _e}} \right){\left( {2\sigma _d^2} \right)^{ - 1}},
\end{align}
}}
and
\begin{align}
{\rm{SOP}}_2^\infty  &= \frac{{m_d^{{m_d}}{d_{{j_d} = 0}}{{\left( {2\sigma _d^2} \right)}^{ - 1}}}}{{\Gamma \left( {{m_d}} \right)}}\left\{ {\frac{{\left( {\lambda  - 1} \right){N_0}}}{{{I_{th}}}}\frac{{m_p^{{m_p}}\left( {2\sigma _p^2} \right)}}{{\Gamma \left( {{m_p}} \right)}}} \right. \notag\\
&\left. {\sum\limits_{{j_p} = 0}^\infty  {\frac{{K_p^{{j_p}}{d_{{j_p}}}\Gamma \left( {{j_p} + 2,\frac{\rho }{{2\sigma _p^2}}} \right)}}{{{j_p}!{j_p}!}}}  + \lambda {\mu _e}{{\overline F}_{{h_p}}}\left( \rho  \right)} \right\},
\end{align}
respectively.

Let ${\left. {{\rm{SOP}}_1^\infty } \right|_{\rho  \to \infty }}$ be the value of ${\rm SOP}_1^\infty$ given $F_{h_p}(\rho)=1$, i.e., $\rho \to \infty$. ${\left. {{\rm{SOP}}_1^\infty } \right|_{\rho  \to \infty }}$ is exactly the ASOP of the three-node Wyner's model investigated in \cite{Zhang2}.

Therefore, the ASOP can be derived as
\begin{align}
{\rm{SO}}{{\rm{P}}^\infty } = {\left( {G2\sigma _d^2} \right)^{ - 1}} + o\left({\left( {2\sigma _d^2} \right)^{ - 2}}\right),
\end{align}
where $G$ is the secrecy array gain, given by
{\color{red}{
\begin{align}
G =& \frac{{\Gamma \left( {{m_d}} \right)}}{{m_d^{{m_d}}{d_{{j_d} = 0}}}}\left\{ {\frac{F_{{h_p}}(\rho)(\lambda-1)N_0}{P_M} + \lambda {\mu _e} + \frac{{\left( {\lambda  - 1} \right){N_0}}}{{{I_{th}}}}} \right. \notag\\
&\cdot \frac{{m_p^{{m_p}}\left( {2\sigma _p^2} \right)}}{{\Gamma \left( {{m_p}} \right)}}{\left. {\sum\nolimits_{{j_p} = 0}^\infty  {\frac{{K_p^{{j_p}}{d_{{j_p}}}}}{{{j_p}!{j_p}!}}} \Gamma \left( {{j_p} + 2,\frac{\rho }{{2\sigma _p^2}}} \right)} \right\}^{ - 1}}.
\end{align}
}}
The expression for ASOP shows that the secrecy diversity order is always unity{\footnote{This unit diversity order conclusion does not include the Nakagami-$m$ fading, a limit special case of FTR fading \cite{Goldsmith}, because some mathematical properties will change if that limit condition happens.}, and the ASOP is a linear function with respect to $2\sigma_d^2$ in the dB scale, where the secrecy diversity order and array gain are the slope and intercept on the abscissa axis, respectively. It is also worth noting that the secrecy array gain ($G$) depends only on the average of the channel power gain of the wiretap channel.
Moreover, by using the relationship between $2\sigma_d^2$ and $\mu_d$, i.e., $\mu_d=2\sigma_d^2(1+K_d)$, we can also obtain the ASOP in terms of $\mu_d$.

\section{Numerical Results}
In calculation of the infinite summation terms in the PDF and CDF of $h_t$ ($t \in \{d,e,p\}$), we truncate the infinite terms into finite terms, where the corresponding truncated error analysis has been evaluated in \cite{Zhang1,Zhang2}. In the analytical results, we truncate the first 80 summation terms from infinite terms, which gives us a very high precision. In the Monte-Carlo simulation, $10^{7}$ channel state realizations are generated to derive the numerical results.

Fig. 1 plots the SOP versus $I_{th}$, where we can easily observe a decreasing trend in SOP with increasing $I_{th}$.  When $I_{th}$ is sufficiently large, the SOP is roughly unchanged, due to the maximal transmit power constraint at $S$, and actually, the SOP can be approximated by ${\rm SOP}_1$ with $\rho \to \infty$, because the cognitive radio (CR) scenario  becomes the non-CR scenario where the transmitter always uses its maximal transmit power. It is obvious that the SOP becomes better as $P_M$ increases, due to the improved transmit power constraint. There is a narrow gap for a larger $P_M$ between the SOP and ${\left. {{\rm{SOP}}_2} \right|_{\rho  = 0}}$, because a larger $P_M$ means a higher probability of ${P_t} = {{{I_{th}}} \mathord{\left/{\vphantom {{{I_{th}}} {{h_p}}}} \right.\kern-\nulldelimiterspace} {{h_p}}}$, which is exactly the power control in CRNs proposed by \cite{Yuanwei2} where the maximal transmit power constraint is not considered.
\begin{figure}[!htb]
\vspace{-0.7cm}  %调整图片与上文的垂直距离
\setlength{\abovecaptionskip}{-0.2cm}   %调整图片标题与图距离
\setlength{\belowcaptionskip}{-3cm}   %调整图片标题与下文距离
\centering
\includegraphics[width= 2 in]{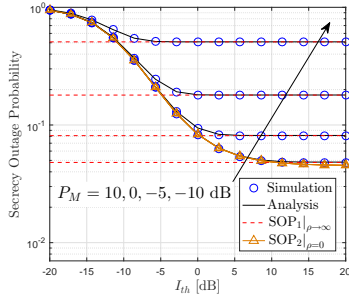}
\caption{Secrecy outage probability versus $I_{th}$ for $K_d=K_e=K_p=10$, $\Delta_d=\Delta_e=\Delta_p=0.5$,    $m_d=m_e=m_p=2.8$, $N_0=0$ dB, $\mu_d=\mu_p=1$, $\mu_e=0.1$, and $R_s=0.1$.}\vspace{-0.4cm}
\label{fig2}
\end{figure}

In Fig. 2, we can see that the SOP becomes better as $\mu_e$ decreases, due to the worse wiretap channel.
The decreasing trend in SOP with respect to $\mu_d$ is shown in Figs. 2-3, where we can also see that the SOP is improved with decreasing  $R_s$ (or increasing $K$), which can be explained by the fact that for a random variable $X$, the probability of $X \le x$ becomes larger for larger $x$ (or the strength of the dominant waves of FTR fading channel grows).
\begin{figure}[!htb]
\vspace{-0.5cm}  %调整图片与上文的垂直距离
\setlength{\abovecaptionskip}{-0.2cm}   %调整图片标题与图距离
\setlength{\belowcaptionskip}{-10cm}   %调整图片标题与下文距离
\centering
\includegraphics[width= 2 in]{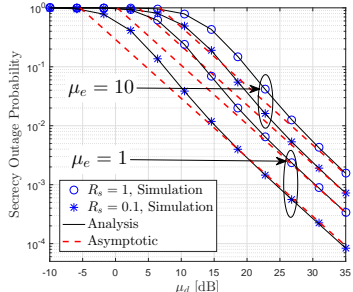}
\caption{Secrecy outage probability versus $\mu_d$ for $K_d=K_e=K_p=5$, $\Delta_d=\Delta_e=\Delta_p=0.5$,    $m_d=m_e=m_p=10.8$, $P_M=N_0=0$ dB, $\rho=-5$ dB, and $\mu_p=1$.}\vspace{-0.5cm}
\label{fig3}
\end{figure}

Further, the slopes  in asymptotic results of Figs. 2-3 are fixed, regardless of any parameter setting, which reflects that the secrecy diversity order is always unity. The impact of all parameters on ASOP is reflected in the intercept on the abscissa axis (i.e., the secrecy array gain).
\begin{figure}[!htb]
\vspace{-0.3cm}  %调整图片与上文的垂直距离
\setlength{\abovecaptionskip}{-0.2cm}   %调整图片标题与图距离
\setlength{\belowcaptionskip}{-10cm}   %调整图片标题与下文距离
\centering
\includegraphics[width= 2 in]{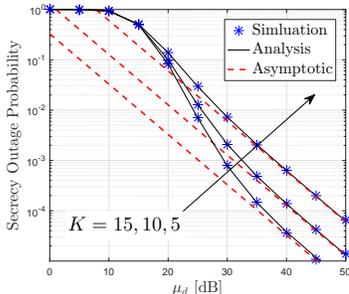}
\caption{Secrecy outage probability versus $\mu_d$ for $K_d=K_e=K_p=K$, $\Delta_d=\Delta_e=\Delta_p=0.5$,    $m_d=m_e=m_p=10.8$, $P_M=N_0=0$ dB, $\rho=-5$ dB, $R_s=3$, and $\mu_e=\mu_p=1$.}\vspace{-0.7cm}
\label{fig4}
\end{figure}

\section{Conclusion}
The analytical expression for the SOP was derived, which can be divided into two parts, i.e., ${\rm SOP}_1$ and ${\rm SOP}_2$. When $\rho = 0$, our SOP becomes the SOP proposed by \cite{Yuanwei2} without taking the maximal transmit power constraint into account. For $\rho \to \infty$, our SOP is reduced to the SOP in non-CRNs investigated in \cite{Zhang2} where the impact of the primary network vanishes. when the SNR of $S-D$ link is sufficiently large, the ASOP shows that the secrecy diversity order is always unity regardless of any parameter setting. From the numerical results, we can conclude that the increase in $\rho$ (or $\mu_d$, $K_d$) and decrease in $\mu_e$ (or $R_s$) will lead to a lower SOP. However, due to the fact that the channel state is uncontrollable, the valid way for the transmitter to improve the SOP is to increase $\rho$ or decrease $R_s$.

\end{document}